\documentclass[12pt]{iopart}

\usepackage{iopams}
\usepackage{graphicx}

\begin{document}

\title[MLDC first round report]{Report on the first round of the Mock LISA Data Challenges}

\author{K A Arnaud$^1$,
G Auger$^2$,
S Babak$^3$,
J Baker$^1$,
M J Benacquista$^4$
E Bloomer$^5$,
D A Brown$^{6,11}$,
J B Camp$^7$,
J K Cannizzo$^7$,
N Christensen$^8$,
J Clark$^5$,
N J Cornish$^9$,
J Crowder$^{9,10}$
C Cutler$^{10,11}$,
L S Finn$^{12}$,
H Halloin$^2$,
K Hayama$^4$,
M Hendry$^5$,
O Jeannin$^2$,
A Kr\'olak$^{13}$,
S L Larson$^{14}$,
I Mandel$^6$,
C Messenger$^5$,
R Meyer$^{15}$,
S Mohanty$^4$,
R Nayak$^4$,
K Numata$^7$,
A Petiteau$^2$,
M Pitkin$^5$,
E Plagnol$^2$,
E K Porter$^{3,9}$,
R Prix$^3$,
C Roever$^{15}$,
B S Sathyaprakash$^{16}$,
A Stroeer$^{17,18}$,
R Thirumalainambi$^{19}$,
D E Thompson$^{19}$,
J Toher$^5$,
R Umstaetter$^{15}$,
M Vallisneri$^{10,11}$,
A Vecchio$^{17,18}$,
J Veitch$^{17}$
J-Y Vinet$^{20}$,
J T Whelan$^3$,
G Woan$^5$}

\address{$^1$ Gravitational Astrophysics Laboratory, NASA Goddard Space Flight Center, 8800 Greenbelt Rd, Greenbelt, MD 20771, US}
\address{$^2$ AstroParticule et Cosmologie (APC), UMR 7164, Universit\'{e} Paris 7 Denis Diderot, 10 rue Alice Domon et L\'{e}onie Duquet, F-75205 Paris Cedex 13, France}
\address{$^3$ Max-Planck-Institut f\"{u}r Gravitationsphysik (Albert-Einstein-Institut), Am M\"{u}hlenberg 1, D-14476 Golm bei Potsdam, Germany}
\address{$^4$ Center for Gravitational Wave Astronomy, University of Texas at Brownsville, Brownsville, TX 78520, US}
\address{$^5$ Dept.\ of Phys.\ and Astron., University of Glasgow, Glasgow G12 8QQ, UK}
\address{$^6$ LIGO Laboratory, California Institute of Technology, Pasadena, CA 91125, US}
\address{$^7$ Laboratory for Gravitational Physics, Goddard Space Flight Center, Greenbelt, MD 20771, US}
\address{$^8$ Phys.\ and Astron., Carleton College, Northfield, MN, US}
\address{$^9$ Dept.\ of Phys., Montana State University, Bozeman, MT 59717, US}
\address{$^{10}$ Jet Propulsion Laboratory, California Institute of Technology, Pasadena, CA 91109, US}
\address{$^{11}$ Theoretical Astrophysics, California Institute of Technology, Pasadena, CA 91125, US}
\address{$^{12}$ Center for Gravitational Wave Physics, The Pennsylvania State University, University Park, PA 16802, US}
\address{$^{13}$ Institute of Mathematics, Polish Academy of Science, Warsaw, Poland}
\address{$^{14}$ Dept.\ of Phys., Weber State University, 2508 University of Circle, Ogden, UT 84408, US}
\address{$^{15}$ Dept.\ of Statistics, The University of Auckland, Auckland, New Zealand}
\address{$^{16}$ Dept.\ of Phys.\ and Astron., Cardiff University, 5, The Parade, Cardiff, CF24 3YB, UK}
\address{$^{17}$ School of Phys.\ and Astron., University of Birmingham, Edgbaston, Birmingham B152TT, UK}
\address{$^{18}$ Dept.\ of Phys.\ and Astron., Northwestern University, Evanston, IL 60208, US}
\address{$^{19}$ NASA Ames Research Center, Moffett Field, CA, US}
\address{$^{20}$ Department ARTEMIS, Observatoire de la C\^{o}te d'Azur, BP 429, 06304 Nice, France}
\ead{benacquista@phys.utb.edu}
\begin{abstract}
The Mock LISA Data Challenges (MLDCs) have the dual purpose of fostering the development of LISA data analysis tools and capabilities, and demonstrating the technical readiness already achieved by the gravitational-wave community in distilling a rich science payoff from the LISA data output. The first round of MLDCs has just been completed: nine challenges consisting of data sets containing simulated gravitational wave signals produced either by galactic binaries or massive black hole binaries embedded in simulated LISA instrumental noise were released in June 2006 with deadline for submission of results at the beginning of December 2006. Ten groups have participated in this first round of challenges. All of the challenges had at least one entry which successfully characterized the signal to better than 95\% when assessed via a correlation with phasing ambiguities accounted for. Here we describe the challenges, summarise the results, and provide a first critical assessment of the entries.

\end{abstract}


\section{Introduction}

At the LISA International Science Team (LIST) meeting of December 2005, the Working Group on Data Analysis (LIST-WG1B) decided to organise several rounds of MLDCs with the dual purposes of (i) fostering the development of LISA data analysis tools and capabilities, and (ii) determining the technical readiness already achieved by the gravitational wave community for distilling a rich science payoff from the LISA data output. These challenges are meant to be blind tests, but not contests. The intent is to encourage the quantitative comparison of results, analysis methods, and implementations.

A MLDC Task Force was constituted at the beginning of 2006 and has been working since then to formulate challenge problems, develop standard models of the LISA mission and gravitational wave (GW) sources, provide computing tools (e.g. LISA response simulators, and source waveform generators), establish criteria for the evaluation of the analyses, and provide any technical support necessary to the challenge participants. The first round of challenges involve the distribution of several data sets, encoded in a simple standard format, and containing combinations of realistic simulated LISA noise with the signals from one or more GW sources with parameters which were unknown to the participants. The participants were then asked to return the maximum amount of information about the sources and to produce technical notes detailing their work.

The release of the first round of challenge data sets was announced in June, 2006 at the Sixth LISA International Symposium hosted by the Goddard Space Flight Center in Greenbelt, Maryland~\cite{MLDCLISA06a, MLDCLISA06b}. John Baker (a member of the MLDC Task Force who did not participate in the first round) was appointed as MLDC1 coordinator. The coordinator was responsible for generating the challenge data sets, receiving the results from the participants, and posting both the key data files and results as soon as possible after the submission deadline of December 4, 2006.

The challenge data sets include a total of 9 year-long data sets which are described in detail on the MLDC website~\cite{MLDCweb}, the Task Force wiki~\cite{MLDCwiki}, and the Omnibus document for Challenge 1~\cite{MLDCdoc}. The challenge data sets are broadly grouped into three categories: (1.1) white dwarf binaries (WDs), (1.2) supermassive black holes (SMBHs), and (1.3) extreme mass ratio inspirals (EMRIs). The problem of detection of EMRIs is considered more difficult than the others, so the deadline for submission of results for the 1.3 challenges is extended to June, 2007. Consequently, in this paper we will discuss the results of challenges 1.1 and 1.2.

The WD challenges consist of three single source data sets with the GW frequency around 1 mHz (1.1.1a), 3 mHz (1.1.1b) and 10 mHz (1.1.1c) and four multiple source data sets with isolated sources of known (1.1.2) and unknown (1.1.3) sky locations and frequencies, and overlapping sources with a low (1.1.4) and high (1.1.5) density of sources in frequency space. The SMBH challenges consist of two single source data sets. In one (1.2.1) the SMBH binary merges during the observation time, and in the other (1.2.2) the merger takes place between one and three months after the end of the data set.

\section{Overview of MLDC1 Submissions}

Ten groups submitted results for MLDC1 by the deadline. These results have been posted on the MLDC website. They include the technical notes submitted by the challenge participants and the files with the ``best parameter fits'' for the data sets. Table~\ref{participation} provides a summary of the groups and their submissions for MLDC1.

\begin{table}
\caption{\label{participation}Groups that participated in the Mock LISA Data Challenge 1. The challenges for which each group submitted results are marked by $\bullet$.}
\begin{indented}
\item[]\begin{tabular}{l|ccccccccc}
\br
Group & \centre{7}{Galactic Binaries} & \centre{2}{Massive} \\
& \centre{3}{Single Source} & \centre{4}{Multiple Sources}& \centre{2}{Black Holes} \\
& \crule{3}&\crule{4}&\crule{2}\\
& 1.1.1a & 1.1.1b & 1.1.1c & 1.1.2 & 1.1.3 & 1.1.4 & 1.1.5 & 1.2.1 & 1.2.2 \\
\mr
AEI & $\bullet$ & $\bullet$ & $\bullet$ & $\bullet$ & $\bullet$ & $\bullet$ & $\bullet$ & & \\
Ames & $\bullet$ & $\bullet$ & & $\bullet$ & $\bullet$ & & & & \\
APC & $\bullet$ & & & & & & & & \\
Goddard & & & & & & & & $\bullet$ & \\
GLIG & & & $\bullet$ & & & & & &  \\
Kr\'olak & $\bullet$ & $\bullet$ & $\bullet$ & & $\bullet$ & & & & \\
JPL/Caltech & & & & & & & & $\bullet$ & \\
MT/AEI & & & & & & & & $\bullet$ & $\bullet$ \\
MT/JPL & $\bullet$ & $\bullet$ & $\bullet$ & $\bullet$ & $\bullet$ & $\bullet$ & $\bullet$ & & \\
UTB & $\bullet$ & $\bullet$ & $\bullet$ & $\bullet$ & $\bullet$ & $\bullet$ & & & \\
\br
\end{tabular}
\end{indented}
\end{table}

With the exception of Challenges 1.1.5 and 1.2.2, every Challenge data set was analysed by at least three groups. Here we briefly summarise the approaches used by each group. More detailed descriptions from many of the groups can be found elsewhere in these proceedings~\cite{auger07,camp07,crowder07,brown07,auger07,nayak07, porter07,prix07, roever07, stroeer07,thompson07} or in the technical notes on the MLDC web page~\cite{MLDCweb}. Several groups used variations on matched filtering methods on many of the challenges. The Ames group at the NASA Ames Research Center employed a user-refined grid search on a number of the WD challenges. The AEI group and Andrzej Kr\'olak both used grid-based methods. The Global LISA Inference Group (GLIG), Montana-JPL, and Montana-AEI groups employed variations on Markov Chain Monte Carlo (MCMC) methods~\cite{gilks96}. The JPL-Caltech group used a multi-stage approach that combined time-frequency methods with grid-based and MCMC searches. The Montana-JPL group also used a genetic algorithm~\cite{holland75}. The APC group has also implemented an hierarchical approach which first matches the annual amplitude modulation and then follows with a full matched filtering. Two groups did not use matched filtering at all. The UTB group used a tomographic search that employed the Radon transform~\cite{deans83} while the Goddard group at the NASA Goddard Space Flight Center developed a time-frequency method that uses the Hilbert-Huang transform~\cite{huang98}. Although some of these groups have well-developed and mature algorithms, most groups are currently in various stages of development and so many of the entries are incomplete or suffer from known bugs which could not be hunted down before the December deadline. Some of the algorithms that are under development are meant to be part of an hierarchical search and so they only return a subset of the parameters needed to fully characterise the source.

\section{Assessment}

The wide variety of approaches and maturity of the algorithms makes it difficult to develop a single assessment that can adequately compare all entries. For those entries that have returned enough parameters to sufficiently generate a recovered waveform, we can compare the recovered waveform $h_{\rm rec}$, with the waveform generated from the ``true'' parameters $h_{\rm key}$, using:
\begin{equation}
\label{deltachi}
\Delta \chi^2 = \frac{\left(h_{\rm key} - h_{\rm rec}|h_{\rm key} - h_{\rm rec}\right)}{D}
\end{equation}
where $(*|*)$ is the noise ($S_n$) weighted inner product summed over channels ($i$), defined by:
\begin{equation}
\left(a|b\right) = 2 \int_{f_{\rm min}}^{f_{\rm max}}{\frac{\sum_{i}{\left(\tilde{a}_ib_i+a_i\tilde{b}_i\right)}}{S_n}df}
\end{equation}
and $D$ is the dimension of the parameter space used to generate the templates. We realise that the $\Delta\chi2$ is not a perfect figure of merit as, for example, it does not account for deduced uncertainties in the recovered parameters. It is however easy to compute and is quite sufficient to indicate whether the recovered parameters differ greatly from those used for the key waveform. The channels used are the noise orthogonal pseudo $A$ and $E$ channels~\cite{prince02}:
\begin{equation}
A = \left(2X - Y - Z\right)/3,~~~~~E = \left(Z - Y\right)/\sqrt{3},
\end{equation}
and $X$, $Y$, and $Z$ are the standard TDI variables. We can also compute the signal-to-noise ratio (SNR) for both $h_{\rm key}$ and $h_{\rm rec}$ using:
\begin{equation}
\label{snr}
{\rm SNR} = \frac{\left(s|h\right)}{\sqrt{\left(h|h\right)}}
\end{equation}
and compare the recovered SNR with the key SNR. Finally, we calculate the correlation between $h_{\rm key}$ and $h_{\rm rec}$ with:
\begin{equation}
\label{cor}
C = \frac{\left(h_{\rm key}|h_{\rm rec}\right)}{\sqrt{\left(h_{\rm key}|h_{\rm key}\right)\left(h_{\rm rec}|h_{\rm rec}\right)}}.
\end{equation}
Some groups reported a known ambiguity in the initial phase and polarisation angles with results being given modulo $\pi/2$ or $\pi$. Obviously, a difference of $\pi$ in the initial phase can significantly degrade the performance of an entry as calculated using $\Delta\chi^2$, SNR, or $C$. Consequently, we have also computed these measures with the initial phase shifted by either $\pi$ or $\pi/2$ as necessary.

Another measure of the success of a given algorithm is the accuracy with which it returns specific parameters. This approach allows us to also evaluate those entries which do not return enough parameters to generate $h_{\rm rec}$. For each parameter $\lambda_i$, we can determine the difference between the key parameter and the recovered parameter using:
\begin{equation}
\Delta\lambda = \lambda_{\rm key} - \lambda_{\rm rec}.
\end{equation}
We note that it is not necessarily appropriate to use the Fisher Information Matrix (FIM) to determine the quality of parameter recovery. If the algorithm settles on a secondary maximum of the likelihood function or there are other systematic errors, then the results can be far from the regime of validity for the FIM approximation to expected errors in parameter estimation. For more details, see Vallisneri's review of the FIM~\cite{vallisneri07}.

The white dwarf binary challenges required the recovery of 7 parameters to fully characterise each source. These parameters are: the amplitude $\mathcal{A}$, the frequency $f$, they sky location $\theta,~\phi$, the angle of inclination $\iota$, the polarisation angle $\psi$, and the initial phase $\phi_0$. In Table~\ref{1.1.1metrics}, we list the values of the measures for each challenge entry for challenges 1.1.1---with the exception of the UTB entry. Since the UTB algorithm only returns frequency (in intervals of resolvable frequency bins $\sim 32$ nHz) and sky position, it cannot be included in this comparison. However, it can be included in the comparison of parameter differences given in Table~\ref{1.1.1parameters}. It should be noted that the GLIG cluster crashed before the completion of the algorithm and therefore the MCMC chain did not have the chance to burn in to the final values. We also note that although the measures of Challenge 1.1.1c in Table~\ref{1.1.1metrics} seem to be quite bad, it is important to note that the accuracy in recovery of the sky positions and frequencies is still comparable to Challenges 1.1.1a and 1.1.1b.

\begin{table}
\caption{\label{1.1.1metrics} The performance of challenge entries on the single binary challenges as calculated using $\Delta\chi^2$, SNR, and $C$. The correction of the initial phase by a factor of $\pi$ or $\pi/2$ is indicated by an asterisk (*).}
\begin{indented}
\item[]\begin{tabular}{lrrr}
\br
Group & $\Delta\chi^2$ & SNR & $C$ \\
\br
\centre{4}{Challenge 1.1.1a (${\rm SNR}_{\rm key}=51.137$)}  \\
\mr
AEI & 8.095 & 50.604 & 0.989\\
Ames & 7.155 & 51.032 & 0.997\\
APC & 423.406 & -8.007 & -0.135 \\
APC* & 229.115 & 50.385 & 0.990 \\
Kr\'olak & 778.888 & 0.933 & -0.004 \\
Kr\'olak* & 1.036 & 51.038 & 0.999 \\
MT/JPL (BAM) & 1.902 & 51.178 & 0.998 \\
MT/JPL (GA) & 1.796 & 51.138 & 0.998 \\
\br
\centre{4}{Challenge 1.1.1b (${\rm SNR}_{\rm key}=37.251$)} \\
\mr
AEI & 47.913 & 33.104 & 0.874\\
Ames & 64.371 & 32.067 & 0.822\\
Kr\'olak & 841.074 & -37.038 & -0.996 \\
Kr\'olak* & 2.566 & 37.038 & 0.996 \\
MT/JPL (BAM) & 7.735 & 36.856 & 0.980 \\
MT/JPL (GA) & 8.371 & 36.808 & 0.979 \\
\br
\centre{4}{Challenge 1.1.1c (${\rm SNR}_{\rm key}=101.390$)} \\
\mr
AEI & 2399.307 & -14.373 & -0.144\\
GLIG & 1788.991 & 14.496 & 0.142\\
Kr\'olak & 5997.595 & -98.126 & -0.968 \\
Kr\'olak* & 97.603 & 98.126 & 0.968 \\
MT/JPL (BAM) & 945.541 & 63.383 & 0.623 \\
MT/JPL (GA) & 1376.143 & 43.564 & 0.424 \\
\br
\end{tabular}
\end{indented}
\end{table}

\begin{table}
\caption{\label{1.1.1parameters} The performance of challenge entries on the single binary challenges as calculated using recovered parameter differences.}
\begin{indented}
\item[]\begin{tabular}{lrrrrrrr}
\br
Group & $\Delta f$ (nHz) & $\Delta \theta$ & $\Delta \phi$ & $\Delta \ln{A}$ & $\Delta \iota$ & $\Delta \psi$ & $\Delta \phi_0$ \\
\br
\centre{4}{Challenge 1.1.1a} \\
\mr
AEI & -1.208 & -0.018 & 0.001 & -0.078 & -0.101 & 0.157 & -0.065 \\
Ames & -1.889 & -1.159 & 3.127 & 0.337 & 0.503 & 0.181 & -0.126 \\
APC & 1.343 & -0.030 & -0.011 & 0.807 & 0.217 & 0.174 & 1.202 \\
Kr\'olak & 0.980 & 0.028 & --0.008 & 0.113 & 0.180 & 0.208 & -2.089 \\
MT/JPL (BAM) & -1.367 & -0.015 & -0.008 & -0.046 & -0.084 & 0.196 & -0.228 \\
MT/JPL (GA) & -1.044 & -0.013 & -0.003 & -0.077 & -0.091 & 0.224 & -0.308 \\
UTB & -3.209 & 0.143 & 0.603 & --- & --- & --- & --- \\
\br
\centre{4}{Challenge 1.1.1b} \\
\mr
AEI & 0.399 & -0.049 & 0.001 & -0.009 & -0.045 & 0.020 & 0.432 \\
Ames & -21.098 & -0.606 & 0.004 & 0.171 & 0.048 & 0.028 & 2.173 \\
Kr\'olak & 0.341 & 0.037 & -0.004 & -0.112 & -0.042 & -0.042 & -3.098 \\
MT/JPL (BAM) & 0.434 & -0.040 & 0.003 & -0.025 & -0.042 & 0.029 & 0.097 \\
MT/JPL (GA) & 0.314 & -0.039 & 0.003 & -0.044 & -0.044 & 0.030 & 0.117 \\
UTB & -4.299 & 0.198 & 0.007 & --- & --- & --- & --- \\
\br
\centre{4}{Challenge 1.1.1c} \\
\mr
AEI & -0.405 & 0.012 & -0.001 & 0.312 & -0.159 & 0.127 & 1.501\\
GLIG & 154.850 & 0.306 & 0.178 & 0.341 & 0.939 & 0.722 & 2.413\\
Kr\'olak & -5.210 & 0.059 & -0.010 & -0.194 & -0.268 & 0.451 & 2.747 \\
MT/JPL (BAM) & -0.330 & 0.008 & -0.001 & 0.309 & 0.033 & -0.609 & 2.148 \\
MT/JPL (GA) & 0.311 & 0.013 & -0.001 & 0.652 & -1.062 & -0.614 & 2.026 \\
UTB & 8.577 & 0.139 & 0.066 & --- & --- & --- & --- \\
\br
\end{tabular}
\end{indented}
\end{table}

The multi-source challenges present a different problem for assessment, since there is the possibility of false positives and false negatives. Consider the possibility in which the recovered parameters for one binary out of many are wildly off. If one were to use the correlation between each recovered template and one of the source binaries, it is possible to count the recovered binary as a false positive and the true binary as a false negative. Given the phasing issues that were apparent in challenges 1.1.1, it is quite likely that there will be several false positive/false negative pairs if such a correlation analysis is used. Consequently, we determine which recovered template goes with which source by looking for template/source pairs that are within one resolvable frequency bin of each other. The overall success of the recovery can be measured using the combined signal of the entire population of recovered binaries as $h_{\rm rec}$ and comparing with the entire population of the true source binaries as $h_{\rm key}$. In this case, we note that the dimension $D$ of the recovered parameter space depends upon the number of recovered sources.

The challenge 1.1.2 data set contained 20 ``verification'' binaries whose frequencies and sky location were given to the participants. Six of these binaries were taken from the list of known binaries available on Gijs Nelemans website~\cite{nelemanswiki} and the remaining 14 were simulated binaries. Of the four groups that submitted entries for challenge 1.1.2, two of them did not use the additional information of sky location and frequency in their searches. The Montana/JPL group used the same search algorithms as they used in all of the 1.1 challenges, and the UTB group used their tomographic algorithm to obtain the frequency and sky location. The UTB group successfully recovered 17 of the 20 source binaries. The other three groups successfully recovered all 20 source binaries, although the AEI group had very low correlation with the 2 highest frequency binaries. This is understandable since the AEI group used the low frequency approximation (which is valid for frequencies below about 3 mHz) for calculating their templates and the 2 highest frequency binaries were at frequencies above 6 mHz. These two binaries are also responsible for most of the loss in correlation for this entry. The performance of the three groups that returned the complete parameterisation of each binary recovered is given in Table~\ref{1.1.2metrics}

\begin{table}
\caption{\label{1.1.2metrics} The performance of challenge entries on the verification binary challenge 1.1.2 as calculated using $\Delta\chi^2$, SNR, and $C$. Since every group returned the full 20 binaries, the dimension of the model is $D = 140$. The SNR of the key is 544.952.}
\begin{indented}
\item[]\begin{tabular}{lrrr}
\br
Group & $\Delta\chi^2$ & SNR & $C$ \\
\mr
AEI & 1443.59 & 339.262 & 0.624\\
Ames & 227.007 & 516.471 & 0.948\\
MT/JPL (BAM) & 19.02 & 544.165 & 0.998 \\
MT/JPL (GA) & 194.46 & 519.712 & 0.954 \\
\br
\end{tabular}
\end{indented}
\end{table}

\begin{figure}
\includegraphics[clip=true,angle=0,width=1.0\textwidth]{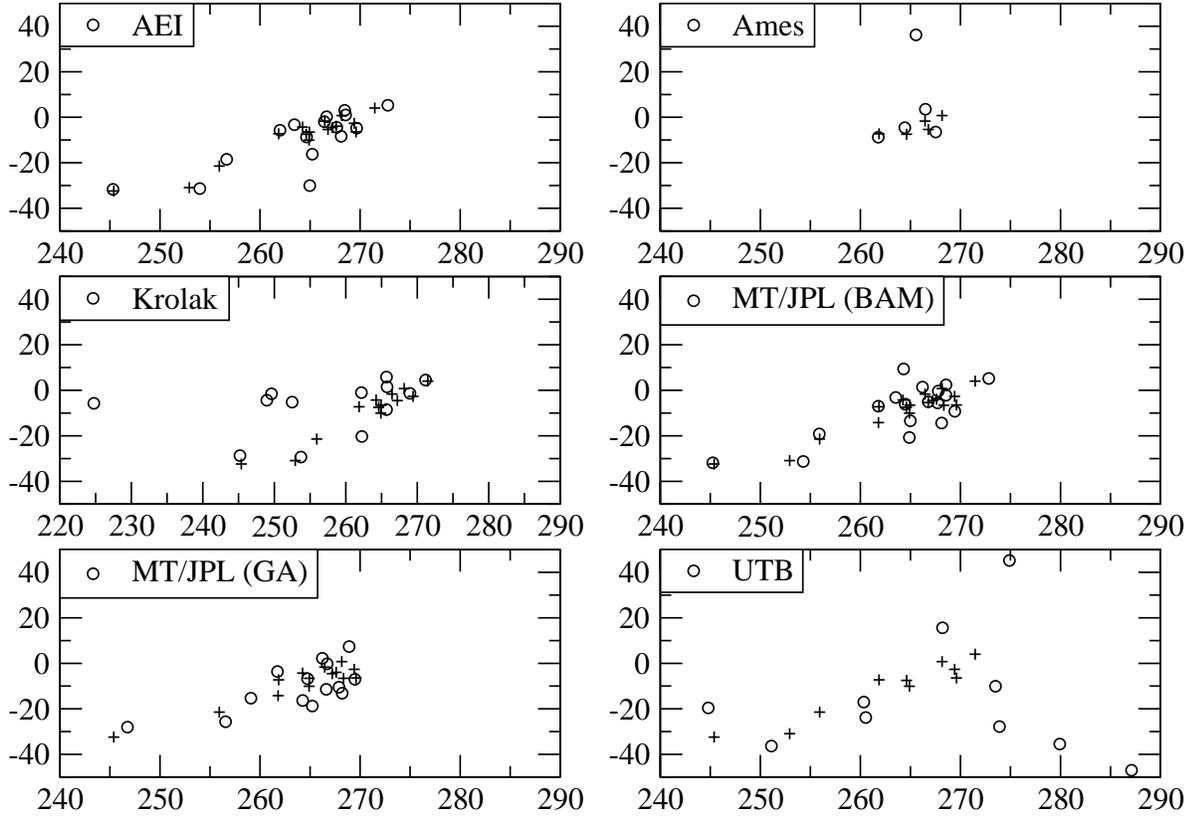}
\caption{\label{1.1.3skypo}Recovered sky positions from each entry for Challenge 1.1.3. The source positions are indicated by $+$ and the recovered positions are indicated by $\circ$. Each plot includes only those sources which are within a frequency bin of a recovered source. The plots are of ecliptic latitude vs. longitude in degrees.}
\end{figure}

Challenge 1.1.3 also contained 20 binaries isolated in frequency space between 0.5 mHz and 10 mHz. In this challenge, all binaries were drawn from the simulation and all parameters were blind. Five groups submitted entries, although the UTB group again only provided frequency and sky location. The UTB group successfully identified 14 of the 20 binaries. The performance of the other four entries are given in Table~\ref{1.1.3metrics}. The sky locations returned by each group are compared with the source positions in Figure~\ref{1.1.3skypo}.

\begin{table}
\caption{\label{1.1.3metrics} The performance of challenge entries on the isolated binary challenge 1.1.3 as calculated using $\Delta\chi^2$, SNR, and $C$. Since not all groups returned the full 20 binaries, the dimension of the model is calculated by $D = 7\times N_{\rm rec}$, where $N_{\rm rec}$ is the number of recovered binaries. The SNR of the key is 122.864. The correction of the initial phase by a factor of $\pi$ or $\pi/2$ is indicated by an asterisk (*).}
\begin{indented}
\item[]\begin{tabular}{lrrrr}
\br
Group & $\Delta\chi^2$ & SNR & $C$ & $N_{\rm rec}$ \\
\mr
AEI & 67.67 & 89.789 & 0.726 & 16\\
Ames & 48.42 & 104.236 & 0.841 & 13\\
Kr\'olak & 204.36 & 9.558 & 0.080 & 20\\
Kr\'olak* & 150.95 & 38.770 & 0.323 & 20\\
MT/JPL (BAM) & 44.81 & 98.838 & 0.797 & 19\\
MT/JPL* (BAM) & 4.16 & 121.64 & 0.981 & 19\\
MT/JPL (GA) & 19.66 & 113.797 & 0.914 & 18\\
\br
\end{tabular}
\end{indented}
\end{table}

The remaining two white dwarf challenges each contained approximately 45 overlapping sources. Challenge 1.1.4 contained 45 sources in a $15 {\rm \mu}$Hz band starting at 3 mHz, while Challenge 1.1.5 contained 33 sources in a $\pm 1.5 {\rm \mu}$Hz band centered on 3 mHz. Challenge 1.1.4 had an average density of 0.095 sources per resolvable frequency bin, and Challenge 1.1.5 had an average density of 0.35 sources per bin. These Challenges were actually more difficult than these source densities might indicate since in both cases, there were at least 3 frequency bins which contained at least 2 binaries each. There was one case in Challenge 1.1.5 with 3 binaries in one frequency bin. Despite this additional complication, both groups managed to recover a respectable number of sources that matched with binaries in the key. The performance of the two groups that submitted complete parameter sets for each binary recovered are listed in Table~\ref{1.1.45metrics}.
\begin{table}
\caption{\label{1.1.45metrics} The performance of challenge entries on the overlapping binary challenges 1.1.4 and 1.1.5 as calculated using $\Delta\chi^2$, SNR, and $C$. Since not all groups returned the full number of binaries, the dimension of the model is calculated by $D = 7\times N_{\rm rec}$, where $N_{\rm rec}$ is the number of recovered binaries. $N_{\rm match}$ is the number of recovered binaries that matched in frequency with a binary in the key.}
\begin{indented}
\item[]\begin{tabular}{lrrrrr}
\br
Group & $\Delta\chi^2$ & SNR & $C$ & $N_{\rm rec}$ & $N_{\rm match}$ \\
\br
\centre{6}{Challenge 1.1.4 (${\rm SNR_{key}} = 201.129$)}\\
\mr
AEI & 85.63 & 159.893 & 0.792 & 26 & 16 \\
MT/JPL (BAM) & 6.19 & 197.828 & 0.976 & 43 & 39 \\
\br
\centre{6}{Challenge 1.1.5 (${\rm SNR_{key}} = 178.261$)}\\
\mr
AEI & 519.21 & 116.822 & 0.654 & 5 & 4 \\
MT/JPL (BAM) & 11.96 & 172.582 & 0.963 & 27 & 23 \\
\br
\end{tabular}
\end{indented}
\end{table}

The supermassive black hole challenges required recovery of 9 parameters describing the source: the chirp mass $\mathcal{M}$, the reduced mass $\mu$, the luminosity distance $D_L$, the time of coalescence $t_c$, the sky location $\theta,~\phi$, the initial angle of inclination $\iota$, the inital polarisation angle $\psi$, and the initial orbital phase $\phi_0$. As with the white dwarf challenges, the quality of the recovered signal can be described by the measures given in Equations~\ref{deltachi}~and~\ref{snr}. In place of the correlation, we compute several overlaps using:
\begin{equation}
\label{overlaps}
O_{\alpha} = \frac{\left(h^{\alpha}_{\rm key}|h^{\alpha}_{\rm rec}\right)}{\sqrt{\left(h^{\alpha}_{\rm key}|h^{\alpha}_{\rm key}\right)\left(h^{\alpha}_{\rm rec}|h^{\alpha}_{\rm rec}\right)}}
\end{equation}
where $\alpha$ denotes the particular TDI channel being used. In order to mitigate the effects of a possible error in the initial phase, we have also computed $O_X$, maximised over the phase:
\begin{equation}
\label{maxphi}
{\rm max}_{\phi_0}\left(O_X\right) = \sqrt{\left(h^{X}_{\rm rec}|h^{X}_{\rm key}(\phi_0=0)\right)^2+\left(h^{X}_{\rm rec}|h^{X}_{\rm key}(\phi_0=\pi/2)\right)^2}.
\end{equation}
There were two groups that returned a full characterisation of the signal for Challenge 1.2.1. The Montana/AEI group had a constant phase difference, and when this phase is corrected, the performance of both the JPL/Caltech and Montana/AEI groups is quite good. The Goddard group is developing a new algorithm using the Hilbert-Huang Transform that is in a very preliminary stage and has only returned $\mathcal{M}$ and $t_c$ for this challenge. Because of a known secondary maximum in sky location, we also check the antipodal sky position:
\begin{equation}
\theta  \rightarrow  - \theta~~~~~,~~~~~\phi  \rightarrow  \phi \pm \pi. \label{skyflip}
\end{equation}
However, this adjustment also requires a change in the values of the inclination and polarisation angles as well. This is accomplished by substituting the initial returned values of $\theta$, $\phi$, $\iota$, and $\psi$ into:
\begin{eqnarray}
\cos{\iota} & = & \cos{\theta}\sin{\Theta}\cos{\left(\phi - \Phi\right)} + \cos{\Theta}\sin{\theta} \label{iota} \\
\tan{\psi} & = & \frac{\sin{\theta}\cos{\left(\phi - \Phi\right)}\sin{\Theta} - \cos{\Theta}\cos{\theta}}{\sin{\Theta}\sin{\left(\phi - \Phi\right)}}, \label{psi},
\end{eqnarray}
and solve these equations for the orientation angles of the orbital angular momentum vector, $\left(\Theta,\Phi\right)$. Once we have the values of $\left(\Theta,\Phi\right)$, we then use these values and the antipodal sky position from Equations~\ref{skyflip} in the above equations~\ref{iota} and~\ref{psi} to determine the new values of $\iota$ and $\psi$. Once these have been found, the new values can be used to provide a more realistic estimate of the error in the returned values. We have applied this transformation to the JPL/Caltech entry and also adjusted the polarisation phase for the Montana/AEI entry in Challenge 1.2.1. Only one group (Montana/AEI) submitted an entry for Challenge 1.2.2. The measures for each submission under both of these challenges are given in Table~\ref{1.2metrics}. We have also determined the errors in the recovered parameters for all entries in Challenges 1.2.1 and 1.2.2. These are presented in Table~\ref{1.2parameters}.
\begin{table}
\caption{\label{1.2metrics} The performance of challenge entries on the supermassive black hole binary challenges 1.2.1 and 1.2.2. The results after a correction of the initial phase are indicated by a *.}
\begin{indented}
\item[]\begin{tabular}{lrrrrr}
\br
Group & $\Delta\chi^2$ & SNR & $O_A$ & $O_E$ & ${\rm max}_{\phi_0}\left(O_X\right)$ \\
\br
\centre{6}{Challenge 1.2.1 (${\rm SNR_{key}} = 667.734$)}\\
\mr
JPL & 261.48 & 664.47 & 0.994 & 0.996 & 0.9955\\
MT/AEI & 10289.29 & 524.29 & 0.790 & 0.791 & 0.9998\\
MT/AEI* & 105.50 & 662.87 & 0.998 & 0.998 & 0.9998\\
\br
\centre{6}{Challenge 1.2.2 (${\rm SNR_{key}} = 104.19$)}\\
\mr
MT/AEI & 1.41 & 104.29 & 0.997 & 0.998 & 0.9955\\
\br
\end{tabular}
\end{indented}
\end{table}

\begin{table}
\caption{\label{1.2parameters} The performance of challenge entries on the supermassive black hole binary challenges 1.2.1 and 1.2.2 on the estimation of recovered parameters. The angles are absolute errors, all other errors are relative.}
\begin{tabular}{lrrrrrrrrr}
\br
Group & $\Delta \mathcal{M}/\mathcal{M}$ & $\Delta\mu/\mu$ & $\Delta D_L/D_L$ & $\Delta t_c/t_c$ & $\Delta\theta$ & $\Delta\phi$ & $\Delta\iota$ & $\Delta\psi$ & $\Delta\phi_0$\\
& $(\times 10^{-4})$ & & & $(\times 10^{-6})$  & & & & \\
\br
\centre{10}{Challenge 1.2.1 (Reported values)}\\
\mr
JPL & 7.35 & 0.011 & 1.101 & 3.35 & 1.030 & -3.170 & 1.32 & -2.65 & 0.004\\
MT/AEI & 0.98 & 0.001 & 0.042 & 0.26 & 0.001 & 0.001 & 0.02 & 3.14 & 0.004\\
Goddard & 434.00 & --- & --- & 113.00 & --- & --- & --- & --- & ---\\
\br
\centre{10}{Challenge 1.2.1 (Angle adjusted values)}\\
\mr
JPL & 7.35 & 0.011 & 1.101 & 3.35 & -0.043 & -0.032 & -0.58 & -0.31 & 0.004\\
MT/AEI & 0.98 & 0.001 & 0.042 & 0.26 & 0.001 & 0.001 & 0.02 & -0.00 & 0.004\\
\br
\centre{10}{Challenge 1.2.2}\\
\mr
MT/AEI & 3.09 & 0.037 & 0.273 & 182.00 & 0.019 & 0.005 & -0.71 & -2.16 & -0.002\\
\br
\end{tabular}
\end{table}

\section{Conclusions}

The first round of the Mock LISA Data Challenges successfully attracted over ten groups to work on the problem of LISA data analysis. These groups attacked several of the challenges with a variety of different approaches. The algorithms and codes used in the challenges were at different levels of maturity and completeness of the pipelines. Nonetheless, all challenges had at least one entry which successfully characterized the signal to better than 95\% when assessed via the correlation with phasing ambiguities accounted for. In the overlapping source challenge 1.1.5, one group was able to recover true binaries at a source density of $\sim 0.25$. Most groups also discovered small bugs or discrepancies in definitions of some of the parameters used to characterise the signal. The first round entries were a success, especially considering that most groups had less than 5 months to work out interfacing issues and assess their codes. Those groups that participated in round 1 have begun implementing the lessons learned for use in round 2~\cite{MLDC2doc}, which is another successful outcome of the challenges. The MLDC Task Force has also begun to address the issues that have been raised as regards assessment of the entries. As the assessments become more refined in future challenges, we anticipate developing assessments for the true LISA data for which there is no key file.

\section*{Acknowledgements}
MB acknowledges funding from NASA Grant NNG04GD52G. MB, SM and RN were supported by the NASA Center for Gravitational Wave Astronomy at University of Texas at Brownsville (NAG5-13396). N Christensen acknowledges funding from NSF Grant PHY-0553422 and the Fulbright Scholar Program. N Cornish, J Crowder and EP acknowledge funding from NASA Grant NNG05GI69G. J Crowder, CC and MV carried out this work at JPL, Caltech, under contract to NASA. MV is grateful for support from the Human Resources Development Fund program at JPL. IM would like to thank the Brinson, Foundation, NASA Grant NNG04GK98G and NSF Grant PHY-0601459 and the LIGO Laboratory. LIGO was constructed by the California Institute of Technology and Massachusetts Institute of Technology with funding from the National Science Foundation and operates under cooperative agreement PHY-0107417.The work of AK was supported in part by grant 1 P03B 029 27 from Polish Ministry of Science and Information Technology. AK would like to acknowledge hospitality of Max Planck Institute for Gravitational Physics in Potsdam, Germany where part of his work was done. AV was partially supported by the Packard Foundation and the NSF.

\end{document}